\documentclass[aps,prl,twocolumn]{revtex4-1}

\usepackage{amsmath}
\usepackage{graphicx}% Include figure files
\usepackage{array}
\usepackage{colortbl}  % For coloring table cells
\usepackage{xcolor}    % For color definitions
\usepackage{dcolumn}% Align table columns on decimal point
\usepackage{bm}% bold math
\usepackage{braket}
\usepackage{nicefrac}
\usepackage[none]{hyphenat}

\usepackage[colorlinks=true,linkcolor=darkblue,urlcolor=darkblue,filecolor=darkblue,citecolor=darkblue,linktocpage=true]{hyperref}

\definecolor{lightgray}{gray}{0.9}
\definecolor{darkblue}{rgb}{0.2, 0, 0.8}

\makeatletter
\def\@fnsymbol#1{$\, \flat$}
\makeatother

\renewcommand{\[}{\begin{equation}\begin{aligned}}
\renewcommand{\]}{\end{aligned}\end{equation}}

\def\pa{\partial}
\def\parad{\partial_r}
\def\patheta{\partial_{\theta}}
\def\pax{\partial_x}

\def\Lz{L_0}

\def\gammaf{\Gamma_4}
\def\cas{\mathcal{C}_2}

\def\so{SO(1,2)}
\def\rp{r_{+}}
\def\rm{r_{-}}
\def\rpm{r_{\pm}}

\def\Zout{{}_s B_{\ell m}^{\text{ref}}}
\def\Zin{{}_s B_{\ell m}^{\text{inc}}}

\def\Rfar{R_{\text{far}}}

\def\gc{g_c}
\def\betac{\beta_c}

\def\Cp{C_+}
\def\Cm{C_-}
\def\Cpm{C_{\pm}}

\def\fp{f_+(x)}
\def\fm{f_-(x)}
\def\fpm{f_{\pm}(x)}
\def\ap{a_+(\tau,\eta)}
\def\am{a_-(\tau,\eta)}
\def\apm{a_{\pm}(\tau,\eta)}
\def\bp{b_+(\tau,\eta)}
\def\bm{b_-(\tau,\eta)}
\def\bpm{b_{\pm}(\tau,\eta)}

\def\fpmo[#1]{f_{\pm}^{(#1)}(x)}
\def\apo[#1]{a_{+}^{(#1)}}
\def\amo[#1]{a_{-}^{(#1)}}
\def\apmo[#1]{a_{\pm}^{(#1)}}
\def\bpo[#1]{b_{+}^{(#1)}}
\def\bmo[#1]{b_{-}^{(#1)}}
\def\bpmo[#1]{b_{\pm}^{(#1)}}

\def\apQCR{a_{+,c}(\tau)}
\def\amQCR{a_{-,c}(\tau)}
\def\apmQCR{a_{\pm,c}(\tau)}
\def\bpQCR{b_{+,c}(\tau)}
\def\bmQCR{b_{-,c}(\tau)}
\def\bpmQCR{b_{\pm,c}(\tau)}

\def\Vqcr{V_{\text{QCR}}}
\def\chiQCR{\chi_{\text{QCR}}}

\begin{document}

	\title{Quantum Criticality in Black Hole Scattering}
	\author{Uri Kol}%
	\email{urikol@fas.harvard.edu}
	\affiliation{%
		Center of Mathematical Sciences and Applications, Harvard University, MA 02138, USA\\
	}%
	
	\date{\today}% It is always \today, today,
	%  but any date may be explicitly specified
	
	\begin{abstract}
The Teukolsky equation describing scattering from Kerr black holes captures a few important effects in the process of binary mergers, such as tidal deformations and the decay of ringdown modes, thereby raising interest in the structure of its solutions. In this letter we identify critical phenomena emerging in the corresponding phase space. One special point exists in this phase space, where the black hole is extremal and the scattered wave lies exactly at the superradiant bound, at which the physics simplifies considerably. We provide an indirect realization of a conformal symmetry emerging at this configuration, which leads to its interpretation as a critical point. Away from the critical point conformal symmetry is broken, but it is shown that critical fluctuations continue to be dominant in a wide range of parameters and at finite black hole temperatures. As in quantum many-body systems, the physics in this regime is described exclusively by the temperature and a set of critical exponents, therefore leading to robust predictions that are unique to the Kerr metric.
	\end{abstract}
	
	%\keywords{Suggested keywords}%Use showkeys class option if keyword
	%display desired
	\maketitle
	
	%\tableofcontents

%%%%%%%%%%%%%%%%%%%%%%%%%%%%%%%%%%%%%%%%%%%%%%%%%%%%%%%%%%%%%%%%
%%%%%%%%%%%%%%%%%%%%%%%%%%%%%%%%%%%%%%%%%%%%%%%%%%%%%%%%%%%%%%%%
\section{Introduction}
%%%%%%%%%%%%%%%%%%%%%%%%%%%%%%%%%%%%%%%%%%%%%%%%%%%%%%%%%%%%%%%%
%%%%%%%%%%%%%%%%%%%%%%%%%%%%%%%%%%%%%%%%%%%%%%%%%%%%%%%%%%%%%%%%

Over the past decade, a number of remarkable discoveries in black hole physics have been made. Since the first direct detection of gravitational waves emitted from a compact binary merger in 2015 \cite{LIGOScientific:2016lio}, the LIGO-Virgo-KAGRA collaboration has cataloged about a hundred events.
On a different front, in 2019 the EHT telescope captured the first-ever image of a black hole in the center of the M87 galaxy \cite{EventHorizonTelescope:2019dse}, and a next-generation experiment is already underway \cite{Ayzenberg:2023hfw}.
The immense technological and experimental progress has led to a strong theoretical effort aiming to produce robust and qualitative predictions for black holes, a goal which remains a challenging problem despite a rich scientific history.

Identifying \emph{symmetry breaking patterns} has proven to be a very useful paradigm in addressing similar challenges in almost all areas of physics. Spontaneous symmetry breaking in particle physics and critical phenomena in condensed matter physics are some of the most well-known manifestations of this paradigm. In black hole physics, patterns of symmetry breaking have been studied with respect to "regional" spacetime symmetries such as the near-horizon symmetries of (near-)extreme Kerr \cite{Bardeen:1999px,Guica:2008mu} and the near-ring symmetries associated with the black hole's photon ring \cite{Hadar:2022xag,Kapec:2022dvc}, as well as hidden symmetries arising in different energy limits \cite{Maldacena:1997ih,Castro:2010fd,Charalambous:2021kcz,Charalambous:2022rre,Hui:2021vcv,Hui:2022vbh}. In this letter we would like to report on a different type of symmetry breaking pattern that occurs in the space of configurations.

Of particular interest in this context is the Teukolsky equation, which describes linear perturbations over the Kerr metric with mass $M$ and angular momentum parameter $a$ \cite{Teukolsky:1973ha}. The Teukolsky equation captures a variety of effects involved in the dynamics of rotating black holes. Some examples include tidal effects during the inspiral phase of a binary merger and quasi-normal mode decay in its ringdown phase. In the eikonal limit, the Teukolsky equation also captures the physics of the photon ring.

\begin{figure}[t]
	%\centering
	\quad\quad\quad
	\includegraphics[width=0.7\textwidth, trim={16cm 10cm 15cm 8cm},clip]{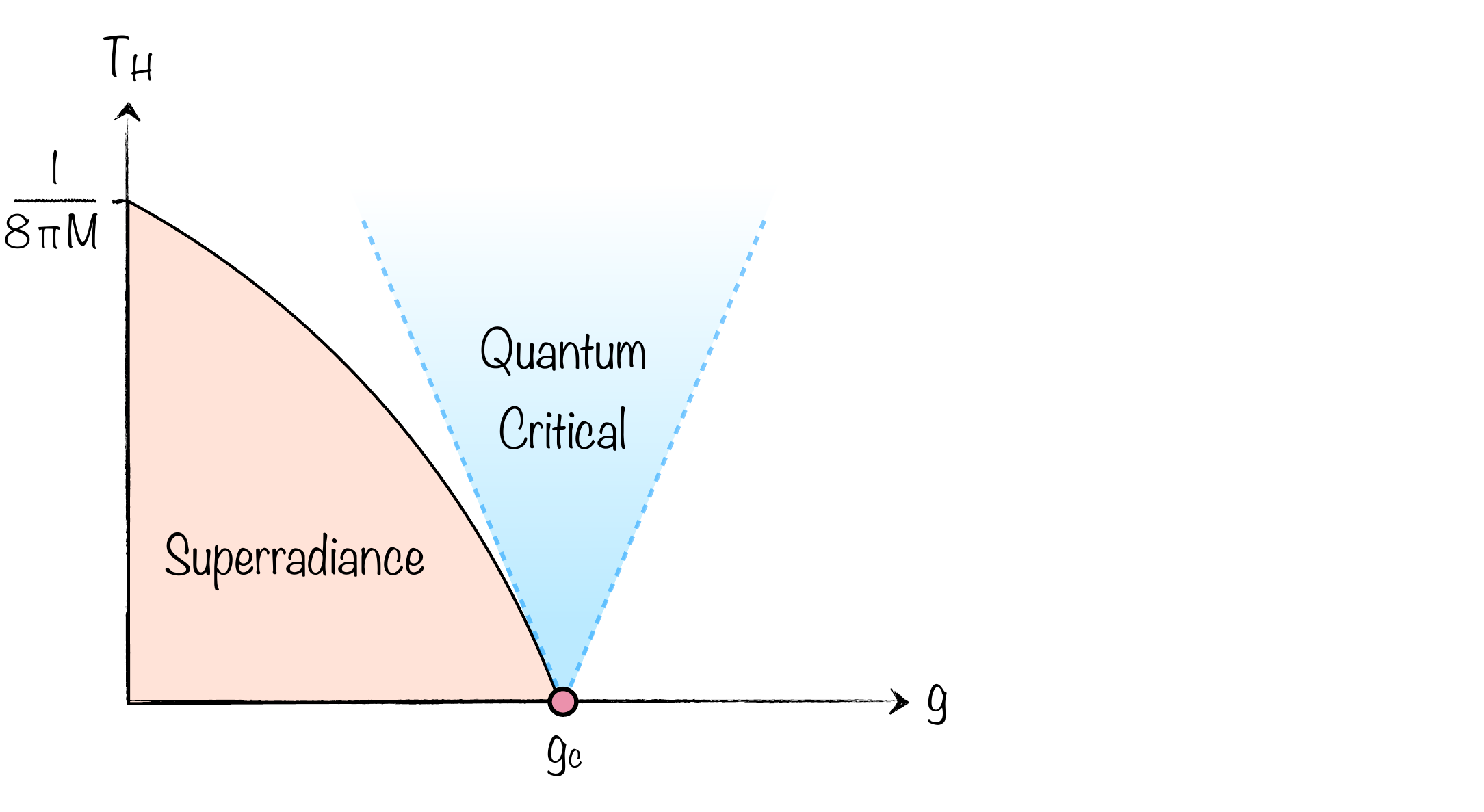} 
	\vspace{-0.8cm}
	\caption{The phase diagram of perturbations around Kerr black holes.}
	\vspace{-0.3cm}
	\label{fig:QCR}
\end{figure}

In the two-dimensional space of configurations spanned by the Hawking temperature and the parameter
\[\label{gparameter}
g \equiv a \omega,
\]
there is a special point where the black hole is extremal and $g$ is equal to
\[
\gc \equiv \frac{m}{2},
\]
corresponding to the superradiant bound. Here $\omega$ and $m$ are the energy and azimuthal number of the perturbation. At this point the radial part of the Teukolsky equation simplifies considerably as it reduces to the confluent hypergeometric equation \cite{Starobinsky:1973aij,starobinskii1973amplification,Teukolsky:1973ha,Press:1973zz,Teukolsky:1974yv,Bredberg:2009pv,Hartman:2009nz}. We provide an indirect realization of a conformal $\so$ symmetry arising in this configuration. The emergence of conformal symmetry at zero temperature is reminiscent of the \emph{Quantum Critical Point (QCP)} in quantum many-body systems \cite{Sachdev:2011fcc,Hartnoll:2018xxg}. At non-zero temperatures  conformal symmetry is broken. However, the scaling behavior implied by the critical point gives rise to a wide range at finite temperatures in which the physics is dominated by critical fluctuations. This region is known in condensed matter physics as the \emph{Quantum Critical Regime (QCR)}. We depict the phase diagram describing these phenomena in figure \ref{fig:QCR}.

%%%%%%%%%%%%%%%%%%%%%%%%%%%%%%%%%%%%%%%%%%%%%%%%%%%%%%%%%%%%%%%%
%%%%%%%%%%%%%%%%%%%%%%%%%%%%%%%%%%%%%%%%%%%%%%%%%%%%%%%%%%%%%%%%
\section{Black Hole Scattering}
%%%%%%%%%%%%%%%%%%%%%%%%%%%%%%%%%%%%%%%%%%%%%%%%%%%%%%%%%%%%%%%%
%%%%%%%%%%%%%%%%%%%%%%%%%%%%%%%%%%%%%%%%%%%%%%%%%%%%%%%%%%%%%%%%

Scattering from Kerr black holes is described by the Teukolsky master equation for linear, spin-$s$, perturbations, which takes the following separable form
\[
\psi = e^{-i \omega t} e^{+i m \phi } S(\theta)R(r),
\]
where the radial and angular equations are given respectively by
\[
&\Delta ^{-s} \parad \left(\Delta^{s+1}\parad R\right) +
\\
& \left(\frac{K^2-2is(r-M)K}{\Delta}+4is\omega r - \lambda\right)R =0,
\]
and
\[
&\frac{1}{\sin \theta} \patheta \left( \sin \theta \patheta S\right)
+\Big(A+
(a\omega)^2 \cos^2 \theta - \frac{m^2}{\sin^2 \theta} 
\\ & \quad
-2a\omega s \cos\theta
-\frac{2ms \cos\theta}{\sin^2\theta}
-s^2 \cot^2 \theta +s 
\Big)S=0.
\]
Here $\Delta = r^2-2Mr+a^2$, $K = (r^2+a^2)\omega -a m$ and $\lambda = A+ (a\omega)^2-2m a \omega$.
The angular wave equation, together with the boundary conditions of regularity at $\theta=0,\pi$, constitutes a Sturm-Liouville eigenvalue problem for the separation constants
\[\label{TeukolskySC}
A={}_sA_{\ell m}(a\omega).
\]
The eigenfunctions of the angular wave equation ${}_s S_{\ell m}(\theta; a \omega)$ are called the \emph{spin-weighted spheroidal harmonics}.

Observables are computed using the asymptotic form of the wavefunction derived from the Teukolsky equation
\[\label{RadAsy}
R(r)=
{}_s R_{\ell m}(r)=
\Zin \, \frac{e^{-i \omega r_{\star}}}{r}
+
\Zout \, \frac{e^{+i \omega r_{\star}}}{r^{1+2s}},
\]
where $r_{\star}$ is the tortoise coordinate. The first term above describes the incoming wave while the second term is the wave reflected from the black hole. The ratio of the reflected to incident amplitudes $\nicefrac{\Zout}{\Zin}$ determines the scattering amplitude and is the main observable of interest here \cite{matzner1978scattering,Futterman:1988ni,Dolan:2008kf,Bautista:2021wfy,Bautista:2022wjf,Ivanov:2022qqt,Saketh:2023bul,Chen:2023qzo,Bautista:2023sdf,Aoki:2024boe,Bautista:2024emt,Kol:2024}. The scattering amplitude computes, in turn, tidal effects in binary mergers \cite{Ivanov:2022qqt,Saketh:2023bul}.

Let us highlight the dependence of the Teukolsky equation on two parameters - the wave's energy $\omega$ and the angular momentum parameter $a$, which can be traded for the Hawking temperature of the black hole
\[
T_H = \frac{1}{8\pi M} \frac{\rp-\rm}{\rp},
\]
where $\rpm=M\pm\sqrt{M^2-a^2}$. We would like to emphasize that while the radial equation depends on both of these parameters explicitly, the angular equation, and in particular the separation constants \eqref{TeukolskySC}, depend only on the combination \eqref{gparameter}. This property of the angular equation will play a key role in our analysis. In particular, the following function of the separation constants will acquire a special meaning
\[
\beta (g) \equiv \sqrt{A(g)+g^2-2m^2 +s(s+1)+\frac{1}{4}}.
\]

When the energy of the incident wave lies in the region
\[
0 < \omega <  m \Omega_H,
\]
where $\Omega_H$ is the angular velocity of the horizon, the reflected wave is amplified with respect to the incident one \cite{Starobinsky:1973aij,starobinskii1973amplification,Teukolsky:1973ha,Press:1973zz,Teukolsky:1974yv}. This phenomena is known as \emph{black hole superradiance}. Exactly at the superradiant bound the absorption cross section vanishes and the wave is perfectly reflected. In figure \ref{fig:QCR}, the superradiant regime is depicted in pink. Note that there is a sharp phase transition between the superradiant phase and the standard scattering phase, which resembles second-order phase transitions in condenses matter systems \cite{Brito:2015oca}. Furthermore, the superradiant phase is reminiscent of long-range ordered phases in quantum many-body systems, as both describe collective phenomena.

In the following, it will be useful to present the results in terms of a redefined radial coordinate
\[
x \equiv \frac{r-\rp}{\rp}.
\]

%%%%%%%%%%%%%%%%%%%%%%%%%%%%%%%%%%%%%%%%%%%%%%%%%%%%%%%%%%%%%%%%
%%%%%%%%%%%%%%%%%%%%%%%%%%%%%%%%%%%%%%%%%%%%%%%%%%%%%%%%%%%%%%%%
\section{The Critical Point}
%%%%%%%%%%%%%%%%%%%%%%%%%%%%%%%%%%%%%%%%%%%%%%%%%%%%%%%%%%%%%%%%
%%%%%%%%%%%%%%%%%%%%%%%%%%%%%%%%%%%%%%%%%%%%%%%%%%%%%%%%%%%%%%%%

The radial Teukolsky equation simplifies dramatically at the critical point, where $T_H=0$ and $g=\gc$, and its solution is given in terms of the confluent hypergeometric function \cite{Starobinsky:1973aij,starobinskii1973amplification,Teukolsky:1973ha,Press:1973zz,Teukolsky:1974yv}
\[\label{criticalSol}
R(x) &= e^{-i \gc x}x^{\betac -s-\frac{1}{2}} 
\times
\\
& {}_1F_1\left(2i \gc+\betac-s+\frac{1}{2};1+2\betac;2i\gc x\right).
\]
Here we refer to the separation constant at the critical point as
\[
\betac \equiv \beta(\gc).
\]

Let us now define the following set of operators
\[
\Lz &\equiv -\frac{x}{2i\gc}  \left(
\frac{\pax x^{2+2s}\pax}{x^{2+2s}}
-\frac{q}{x^2}+\gc^2
\right),\\
\gammaf &\equiv -\frac{x}{2i\gc}  \left(
\frac{\pax x^{2+2s}\pax}{x^{2+2s}}
-\frac{q}{x^2}-\gc^2
\right),\\
D &\equiv -i (x \pax +s+1),
\]
where $q=\betac^2-\frac{1}{4}-s(s+1)$.
This set obeys the $\so$ algebra
\[
\left[ \Lz,\gammaf \right] = +i D,
\,\,\,
\left[ \gammaf , D \right] = -i \Lz,
\,\,\,
\left[  D, \Lz  \right] = +i \gammaf,
\]
with the Casimir operator given by
\[
\cas=\Lz^2 -\gammaf^2-D^2 
=  \betac^2 - \frac{1}{4} .
\]

The radial Teukolsky equation at the critical point can be written as
\[\label{radTeukolskyOp}
\Lz R(x) = -( 2 i \gc -s) R(x).
\]
Note that $\Lz$ is not a Hermitian operator and therefore solutions of \eqref{radTeukolskyOp} do not furnish unitary representations of $\so$. The algebra defined above might seem useless at this point. However, it is possible to associate meaning to it by analytically continuing the spacetime coordinates $t$ and $\phi$. Under this analytic continuation $\gc \rightarrow - i \gc$ and therefore both operators $\Lz$ and $\gammaf$ become Hermitian. In this case, solutions of the radial Teukolsky equation do sit in unitary irreducible representations of $\so$ with real - and quantized - eigenvalues, and the action of the operators $\gammaf,D$ maps them into each other. This symmetry is a remnant of the much larger symmetry group of the self-dual black hole \cite{Guevara:2023wlr}. For more details see \cite{Kol:2024}.
Let us emphasize that we are not interested here in the analytically continued spacetime beyond the argument made above, which allows us to interpret this configuration as a critical point. Ultimately, realistic configurations constitute of rotating black holes at finite temperatures, in which case the conformal symmetry is broken anyway and in any signature. However, as we will see in the next sections, the symmetry breaking pattern at finite temperatures constrains the dynamics \emph{even in Lorentzian signature} and we will use it to compute observables that probe astrophysical Kerr black holes.

%%%%%%%%%%%%%%%%%%%%%%%%%%%%%%%%%%%%%%%%%%%%%%%%%%%%%%%%%%%%%%%%
%%%%%%%%%%%%%%%%%%%%%%%%%%%%%%%%%%%%%%%%%%%%%%%%%%%%%%%%%%%%%%%%
\section{Scaling Limit}
%%%%%%%%%%%%%%%%%%%%%%%%%%%%%%%%%%%%%%%%%%%%%%%%%%%%%%%%%%%%%%%%
%%%%%%%%%%%%%%%%%%%%%%%%%%%%%%%%%%%%%%%%%%%%%%%%%%%%%%%%%%%%%%%%

In the vicinity of the critical point we consider the following scaling limit
\[\label{ScalingLimit}
\tau \equiv 8 \pi M \times T_H\ll 1  \quad
\text{with} \quad 
\eta \equiv \frac{g-\gc}{\tau} \quad \text{fixed}.
\]
This limit was studied in \cite{Teukolsky:1973ha,Press:1973zz,Teukolsky:1974yv,Starobinsky:1973aij,starobinskii1973amplification,Bredberg:2009pv,Hartman:2009nz} to leading order in $\tau$ and it is essentially a consequence of the conformal symmetry discussed in the previous section. In this limit one can solve for the radial wavefunction perturbatively in $\tau$. The solution is constructed by solving separately in the near-horizon region $x\ll 1$, where boundary conditions are imposed, and in the far region $x\gg \tau$, where observables are defined. By matching the two solutions in the region where they overlap $\tau \ll x \ll1$ one can determine the asymptotic form of the solution and compute the scattering amplitude. In the following we will adopt the strategy of \cite{Faulkner:2009wj} for constructing the solution order by order in $\tau$.

In the far region, the two linearly independent solutions can be parametrized as follows
\[
\fpm = \fpmo[0] + \tau  \, \fpmo[1] + \tau^2 \, \fpmo[2]+\dots  
\]
and the wavefunction is a linear combination of them
\[
\Rfar(x) = \Cp \, \fp + \Cm \, \fm .
\]
The functions $\fpmo[n]$ have the following asymptotic behavior
\[
\fpmo[n] =
\bpmo[n] \, \frac{e^{-i \omega x_{\star}}}{x}
+
 \apmo[n] \, \frac{e^{i \omega x_{\star}}}{x^{2s+1}}
\]
at large $x$, with coefficients $\apmo[n],\bpmo[n]$ to be computed.
Comparing to \eqref{RadAsy}, we can express the reflected and incident waves' amplitudes as 
\[
\Zout &= \ap  \, \Cp +\am \, \Cm \, ,
\\
\Zin &= \bp \, \Cp + \bm \, \Cm \, ,
\]
where
\[\label{tauExpansion}
\apm &= \apmo[0] + \tau \, \apmo[1](\eta) + \tau^2 \, \apmo[2](\eta)+\dots ,
\\
\bpm &= \bpmo[0] + \tau \,  \bpmo[1](\eta) + \tau^2 \, \bpmo[2](\eta)+\dots.
\]
Note that we have highlighted the dependence on the parameters $\tau$ and $\eta$, and suppressed the dependence on the quantum numbers $\ell,m$ and $s$.

So far we have only presented the general form of the radial wavefunction. To fully determine the solution one needs to do two things. First, the connection coefficients $\Cpm$ are computed by fixing boundary conditions at the horizon and transferring this information to the far region using the matching procedure. The result is given by %\cite{Starobinsky:1973aij,starobinskii1973amplification,Teukolsky:1973ha,Press:1973zz,Teukolsky:1974yv,Bredberg:2009pv,Hartman:2009nz}
\[\label{susceptibility}
&\chi \equiv \frac{\Cp}{\Cm} = \tau^{-2\betac} \times \\ 
&\quad \frac{
	\Gamma (+2\betac)\Gamma (\frac{1}{2}-\betac-s-im)\Gamma (\frac{1}{2}-\betac-2im\eta)
}{
	\Gamma (-2\betac)
	\Gamma (\frac{1}{2}+\betac-s-im)\Gamma (\frac{1}{2}+\betac-2im\eta)
}
\]
\cite{Starobinsky:1973aij,starobinskii1973amplification,Teukolsky:1973ha,Press:1973zz,Teukolsky:1974yv,Bredberg:2009pv,Hartman:2009nz}.
In \cite{Bredberg:2009pv,Hartman:2009nz}, $\chi^{-1}$ was interpreted as the retarded Green's function in a dual CFT description of the near-horizon region. In condensed matter language $\chi$ is the susceptibility. At the critical point $\tau\rightarrow 0$ and the susceptibility diverges with a rate determined by $\betac$, which is naturally interpreted as the critical exponent. Note that $\betac$ silently carries the indices $\ell,m$ and $s$ and therefore every such mode has its own critical exponent.  We present several examples for the values of the critical exponents $\betac$ in table \eqref{table:criticalExp}. The key point that was emphasized in \cite{Faulkner:2009wj}, albeit for a different system, is that the connection coefficients $\Cpm$ are computed once and for all using the leading order computation in $\tau$. This is a consequence of the linearity of the perturbation equation, which implies that higher orders in $\tau$ will not alter this result.

The second task is to compute the functions $\apmo[n](\eta)$ and $\bpmo[n](\eta)$ order by order in the far region. The leading order in $\tau$, far region, solutions are given by
\[
\fpmo[0] &= e^{-\frac{imx}{2}} \, x^{-\frac{1}{2} \pm \betac-s} \times 
\\
&
 {}_1F_1\left(\frac{1}{2} \pm \betac-s + i m;1\pm 2\betac;imx\right)
\]
and the respective coefficients of their asymptotic expansion are computed to be
\[
a_{\pm}^{(0)} &= \left(+i m \right)^{i m -\frac{1}{2}\mp\betac-s}
\frac{\Gamma(1\pm2\betac)}{\Gamma(\frac{1}{2} \pm \betac-s+im)},
\\
b_{\pm}^{(0)} &= \left(-i m \right)^{-i m -\frac{1}{2}\mp\betac+s}
\frac{\Gamma(1\pm2\betac)}{\Gamma(\frac{1}{2} \pm \betac+s-im)}.
\]
Higher order coefficients in the asymptotic expansion of the far region solutions can be computed numerically.

\begin{table}[t]
	\centering
	\begin{tabular}{>{\centering\columncolor{lightgray}$}p{1cm}<{$} >{$}c<{$} >{$}c<{$} >{$}c<{$} >{$}c<{$} >{$}c<{$}}  % Use `>{$}c<{$}` for math mode
		m& \ell=m & \ell=m+1 & \ell=m+2 & \ell=m+3 & \ell=m+4  \\
		\hline
		\rule{0pt}{1.2em} 
	0 & - & - & \nicefrac{5}{2}  & \nicefrac{7}{2}  & \nicefrac{9}{2} \\
1 & - & 1.919 & 3.175 & 4.267 & 5.317 \\
2 & 2.051 i & 1.871 & 3.480 & 4.728 & 5.872 \\
3 & 2.794 i & 1.376 & 3.543 & 4.985 & 6.246 \\
4 & 3.565 i & 1.070 i & 3.387 & 5.075 & 6.474 \\
	\end{tabular}
	\caption{Several examples of the $\betac$ values for a graviton perturbation $s=2$, computed using \cite{BHPToolkit}. Recall that the various quantum numbers take the values $\ell=|s|,|s|+1,\dots$ and $-\ell \le m \le +\ell$. Note that $\betac$ can be real or imaginary \cite{Bardeen:1999px,Bredberg:2009pv}. Explicit values of the critical exponents for scalar and vector perturbation are considered in \cite{Kol:2024}.}
	\label{table:criticalExp}
\end{table}

To summarize, the problem is factorized into two parts: the connection coefficients $\Cpm$ carry the information about boundary conditions at the horizon, while the functions $\apmo[n](\eta)$ and $\bpmo[n](\eta)$ are solely determined by the kinematics of the far region.

Finally we arrive at the expression for the ratio
\[\label{formalSol}
\frac{\Zout}{\Zin} = \frac{ \ap +\am \, \chi }{\bp+\bm \, \chi } 
\]
which determines the scattering amplitude. Beyond the leading order in $\tau$, this expression for the scattering amplitude is still too formal. In the next section we will see that dramatic simplification occurs in the quantum critical regime.

%%%%%%%%%%%%%%%%%%%%%%%%%%%%%%%%%%%%%%%%%%%%%%%%%%%%%%%%%%%%%%%%
%%%%%%%%%%%%%%%%%%%%%%%%%%%%%%%%%%%%%%%%%%%%%%%%%%%%%%%%%%%%%%%%
\section{Quantum Critical Regime}
%%%%%%%%%%%%%%%%%%%%%%%%%%%%%%%%%%%%%%%%%%%%%%%%%%%%%%%%%%%%%%%%
%%%%%%%%%%%%%%%%%%%%%%%%%%%%%%%%%%%%%%%%%%%%%%%%%%%%%%%%%%%%%%%%

A formal solution to the Teukolsky equation was constructed in the previous section as a function of the two parameters, $\tau$ and $\eta$, that span the phase space of perturbations over the Kerr metric. The solution \eqref{formalSol} depends on four kinematical functions $\apm,\bpm$ that can be expressed as a double expansion in $\tau$ and $\eta$. Each of the coefficients in the $\tau$-expansion \eqref{tauExpansion} can be further expanded in $\eta$
\[\label{etaExpansion}
\apmo[n](\eta) &= \apmo[n](0)+ \sum_{k=1}^{\infty}\frac{1}{k!}\eta^k\pa_{\eta}^k\apmo[n](0) ,
\\
\bpmo[n](\eta) &= \bpmo[n](0)+ \sum_{k=1}^{\infty}\frac{1}{k!}\eta^k\pa_{\eta}^k\bpmo[n](0) .
\]
Derivative terms in the $\eta$-expansion will depend on derivatives of the function $\beta(g)$ evaluated at $g=\gc$
\[
\pa_{\eta}^k\apmo[n](0),\pa_{\eta}^k\bpmo[n](0) \quad \sim \quad  \beta^{(k)}(\gc)+\dots ,
\]
where the dots stand for dependencies on lower order derivatives of $\beta(g)$. However, the leading order coefficients in the $\eta$-expansion \eqref{etaExpansion} depend on $\betac$ only and not on any derivatives of $\beta(g)$. We therefore conclude that to leading order in the approximation
\[\label{QCregime}
\eta \ll 1,
\]
the kinematic functions appearing in \eqref{formalSol} solely depend on $\tau$ and $\betac$. We refer to the regime \eqref{QCregime} as the \emph{quantum critical regime}.

The scattering amplitude in the critical regime therefore reduces to
\[
\left( \frac{\Zout}{\Zin} \right)_{\text{QCR}}=\frac{ \apQCR +\amQCR \, \chiQCR }{\bpQCR+\bmQCR \,  \chiQCR },
\]
where the susceptibility \eqref{susceptibility} simplifies to
\[\label{QCRsusceptibility}
\chiQCR = 
2^{4\betac}
\frac{
	\Gamma (+\betac)\Gamma (\frac{1}{2}-\betac-s-im)
}{
	\Gamma (-\betac)\Gamma (\frac{1}{2}+\betac-s-im)
}\,\,
\tau^{-2\betac}.
\]
The kinematic functions $\apmQCR,\bpmQCR$ only depend on the critical exponent $\betac$, as indicated by the subscript $c$, and not on any other separation constants appearing in the Teukolsky equation. They are computed using the reduced radial Teukolsky equation in the critical regime
\[
\kern-3.8pt    ( x(x+\tau) \pax ^2 +(1+s)(2x+\tau)\pax + \Vqcr  )   R(x) =0,
\]
where the potential is given by
\[
\Vqcr &= \frac{1}{4}-m^2+s(s+1)-\betac^2
\\
+& \,\,\,\, \frac{2ims(1+x)}{\sqrt{1-\tau}}
+\frac{ H( H-is\sqrt{1-\tau}(2x+\tau))}{(1-\tau)x(x+\tau)},
\\
H&=\frac{m}{2}(x(x+2)+\tau).
\]

As in quantum many-body systems, the crossovers from the quantum critical regime to other phases of the system (represented by dashed blue lines in figure \ref{fig:QCR}), are not sharp phase transitions. We also note that the critical theory of black hole perturbations breaks down when the angular momentum is set to zero, namely at high enough temperatures, since the collective phenomena of superradiance seizes to exist at this point. This feature is reminiscent of the breakdown of the critical theory in condensed matter systems at high temperatures corresponding to the microscopic lattice scale. Finally, let us emphasize that the existence, and structural shape, of the critical regime, crucially depend on the scaling limit \eqref{ScalingLimit} exhibited by the system. Systems with different scaling behaviors will show different critical structures.

%%%%%%%%%%%%%%%%%%%%%%%%%%%%%%%%%%%%%%%%%%%%%%%%%%%%%%%%%%%%%%%%
%%%%%%%%%%%%%%%%%%%%%%%%%%%%%%%%%%%%%%%%%%%%%%%%%%%%%%%%%%%%%%%%
\section{Discussion}
%%%%%%%%%%%%%%%%%%%%%%%%%%%%%%%%%%%%%%%%%%%%%%%%%%%%%%%%%%%%%%%%
%%%%%%%%%%%%%%%%%%%%%%%%%%%%%%%%%%%%%%%%%%%%%%%%%%%%%%%%%%%%%%%%

Black hole perturbation theory greatly simplifies in the quantum critical regime. In particular, the full dependence of the system on the Teukolsky-Starobinsky separation constants is relaxed and the physics is solely determined by the Hawking temperature and a set of critical exponents. However, the importance of the critical regime is not merely in simplifying the description. Critical exponents are typically perceived as a unique characterizing measure and therefore provide a robust signature of the system. The extension of the critical regime to finite temperatures grants, in turn, a realistic opportunity to probe the object, whether a black hole or quantum matter.

%%%%%%%%%%%%%%%%%%%%%%%%%%%%%%%%%%%%%%%%%%%%%%%%%%%%%%%%%%%%%%%%
%%%%%%%%%%%%%%%%%%%%%%%%%%%%%%%%%%%%%%%%%%%%%%%%%%%%%%%%%%%%%%%%
\subsection{Acknowledgments}
%%%%%%%%%%%%%%%%%%%%%%%%%%%%%%%%%%%%%%%%%%%%%%%%%%%%%%%%%%%%%%%%
%%%%%%%%%%%%%%%%%%%%%%%%%%%%%%%%%%%%%%%%%%%%%%%%%%%%%%%%%%%%%%%%

I would like to thank Maria Rodriguez for numerous insightful discussions that have ultimately led to the work presented in this letter.
I would like to thank Shing-Tung Yau, Alfredo Guevara, Huy Tran and Mikhail Ivanov for collaborations on related topics and for comments on the manuscript.
UK is supported by the Center for Mathematical Sciences and Applications at Harvard University.

%%%%%%%%%%%%%%%%%%%%%%%%%%%%%%%%%%%%%%%%%%%%%%%%%%%%%%%%%%%%%%%%
%%%%%%%%%%%%%%%%%%%%%%%%%%%%%%%%%%%%%%%%%%%%%%%%%%%%%%%%%%%%%%%%
%%%%%%											  	Bibliography											  	 %%%%%%
%%%%%%%%%%%%%%%%%%%%%%%%%%%%%%%%%%%%%%%%%%%%%%%%%%%%%%%%%%%%%%%%
%%%%%%%%%%%%%%%%%%%%%%%%%%%%%%%%%%%%%%%%%%%%%%%%%%%%%%%%%%%%%%%%

\bibliography{bibliography}

\end{document}